\makeatletter
\declare@file@substitution{revtex4-1.cls}{revtex4-2.cls}
\makeatother
\documentclass{aastex631}
\usepackage{hyperref}
\usepackage{float}

\begin{document}

\title{Orbital phase shifts of Type-I outbursts in EXO 2030+375}

\author{Run-Ting Huang}
\affiliation{School of Physics and Astronomy, Sun Yat-sen University, Zhuhai, 519082, People's Republic of China}

\author[0000-0001-9599-7285]{Long Ji\textsuperscript{*}}
\email{jilong@mail.sysu.edu.cn}
\affiliation{School of Physics and Astronomy, Sun Yat-sen University, Zhuhai, 519082, People's Republic of China}
\affiliation{CSST Science Center for the Guangdong-Hong Kong-Macau Greater Bay Area, DaXue Road 2, 519082, Zhuhai, People's Republic of China}

% \author[0000-0002-5830-3544]{Pu Du}
% \affiliation{Key Laboratory of Particle Astrophysics, Institute of High Energy Physics, Chinese Academy of Sciences, 19B Yuquan Road, Beijing 100049, China}

\author{Jing-Zhi Yan}
\affiliation{Key Laboratory of Dark Matter and Space Astronomy, Purple Mountain Observatory, Chinese Academy of Sciences, Nanjing 210023, People's Republic of China}



\begin{abstract}
EXO 2030+375 is a peculiar high-mass X-ray binary that has been exhibiting Type-I outburst activities consistently over the past decades. 
The phases of outburst peaks are generally stable and occur near the orbital periastron.
However, significant orbital phase shifts have occasionally been observed in 1995, 2006, and 2016. 
In this paper, we report another orbital phase shift triggered by a giant outburst in 2021. 
During this event, the orbital phase of Type-I outburst peaks changed dramatically, moving from near periastron to 20 days after periastron, followed by a gradual recovery. 
Additionally, for several cycles after the 2021 giant outburst, some Type-I outbursts were missing, with subsequent outbursts occurring every two orbits. 
We discuss our observations in the frame of the precession of an eccentric Be disk.
\end{abstract}



\section{Introduction} \label{sec:intro}
Be X-ray binaries (BeXRBs) consist of a compact object (typically a neutron star) and a main-sequence star of type O9-B2 that shows Balmer emission lines at least once. %and potentially shows emission lines of He and Fe.
The emission lines originate from the circumstellar disk surrounding the Be star.
The circumstellar disk is geometrically thin and Keplerian, formed by viscous diffusion of matter ejected from the equator of the Be star.
It is commonly referred to as a \textit{decretion} disk \citep{Okazaki2001, Rivinius2013}.
%As Be stars rotate fast enough to reach their critical velocity, they create a circumstellar disk which can be described as viscous decretion disks \citep{Rivinius2013}.
%When a neutron star approaches or passes through the decretion disk, it can induce density perturbations in the disk and accrete material from the donor star.
If the radial extent of the decretion disk is large enough to fill in the Roche-lobe, the neutron star will accrete material from the donor, converting the gravitational energy of the accreting material into X-ray radiation near the neutron star, leading to outbursts.
In observations, most BeXRBs are X-ray transient systems, and their outbursts can be further divided into two categories: (i) Type-I outbursts (or normal outbursts), which occur (quasi-)periodically when the neutron star reaches the orbital periastron. 
They generally have a peak luminosity of $L_{\mathrm{x}} \sim 10^{36}-10^{37}$ erg\,s$^{-1}$ and a short duration, i.e., about 0.2-0.3 orbital periods.
(ii) Type-II outbursts (or giant outbursts), which can last for multiple orbits and reach luminosities up to $L_{\mathrm{x}} \geq 10^{37}$ erg\,s$^{-1}$ \citep[see, e.g.,][]{Reig2011, Weng2024}.
Although both of them arise from accretion, their underlying mechanisms are different.
Type-I outbursts generally appear when the neutron star approaches or passes through the Be disk because of either an eccentric binary orbit or an eccentric decretion disk \citep{Okazaki2001, Okazaki2013, Franchini2019}.
However, the physics behind Type-II outbursts remains unclear.
Theoretical studies suggest that they should be associated with a highly misaligned and eccentric decretion disk \citep{Okazaki2013, Martin2014a},
perhaps caused by Kozai-Lidov oscillations \citep{Martin2014b, Fu2015, Laplace2017}.  
Optical observations have strongly supported this idea via variations of $\rm H_{\alpha}$ line profiles \citep{Moritani2013P} and polarization parameters \citep{Reig2018}.

Previous studies have predominantly concentrated on either Type-I or Type-II outbursts.
However, given that both phenomena originate from accretion from the donor, exploring their connection could offer valuable insights into the structure and the physics of the decretion disk surrounding the Be star.
For example, orbital phases of Type-I outbursts underwent a sudden shift from 6 to 13 days after the periastron passage, associated with the 2006 giant outburst in EXO 2030+375 \citep{Baykal2008, Laplace2017}.
A similar phase shift was also observed in 1995 (albeit in the opposite direction, i.e., from 6\, days after periastron to 4\, days before periastron)  when no giant outburst occurred, which was explained as evidence of a global one-armed oscillation in the Be disk \citep{Wilson2002, Wilson2005}.
EXO 2030+375 is a BeXRB composed of a B0 Ve star and a magnetized neutron star with a spin period of 42 seconds \citep{Coe1988, Parmar1989}.
It has an orbital period of $46.0205 \pm 0.0002$ days and an eccentricity of $0.412 \pm 0.001 $ \citep{Wilson2008}. 
The distance is $2.4_{-0.4}^{+0.5}$\,kpc estimated by using \textit{Gaia} \citep{BailerJones2021}.
Since its discovery, three Type-II outbursts have been detected in 1985, 2006, and 2021, respectively \citep[e.g.,][]{Parmar1989, Klochkov2007, Fu2023, Thalhammer2024}.
This source exhibits an atypical behavior among BeXRBs, characterized by sustained Type-I outbursts over several years.
Thus, it is the best target for studying the orbital phase evolution of Type-I outbursts and Type-II outbursts.
In this paper, we aim to investigate the behavior of the recent giant outburst in 2021 and compare it with previous studies.

\section{Observations}

The Burst Alert Telescope (BAT) onboard the {\it Swift} observatory is an all-sky hard X-ray monitor operating in the 15-50\,keV energy band. It is designed to detect transient events and monitor the evolution of persistent sources \citep{Gehrels2004, Barthelmy2005}. 
We used the daily lightcurve of EXO 2030+375 as an indicator of the bolometric flux during outbursts, provided by the Hard X-ray Transient Monitor website\footnote{\url{https://swift.gsfc.nasa.gov/results/transients/EXO2030p375}}.
We also tracked EXO 2030+375's outburst activities at soft X-rays measured by the Gas Slit Camera on board the Monitor of All-Sky X-ray Image\footnote{\url{https://maxi.riken.jp/star_data/J2032+376/J2032+376.html}} \citep[MAXI;][]{Matsuoka2009}.
%\subsection{NEOWISE}

The NEOWISE project was the asteroid-hunting portion of the Wide-field Infrared Survey Explorer (WISE) mission launched in December 2009 \citep{Wright2010, Mainzer2014}. 
It conducted the science survey at W1 (3.4\,$\mu$m) and W2 (4.6\,$\mu$m) bands after the reactivation in 2013 until July 2024.
We used its W1 band lightcurve downloaded from the Infrared Science Archive (IRSA) viewer\footnote{\url{irsa.ipac.caltech.edu/irsaviewer/}}.

%\subsection{ZTF}
The Zwicky Transient Facility (ZTF) is a public-private partnership to systematically study the optical night sky \citep{Bellm2019}. 
ZTF scans the entire Northern sky every two days using an extremely wide-field view camera.
We adopted its long-term lightcurves of g (480\, nm) and r (650 nm) bands from IRSA\footnote{https://irsa.ipac.caltech.edu/Missions/ztf.html}.

\section{Data analysis and results}
\begin{figure*}
    \centering
    \includegraphics[width=1\linewidth]{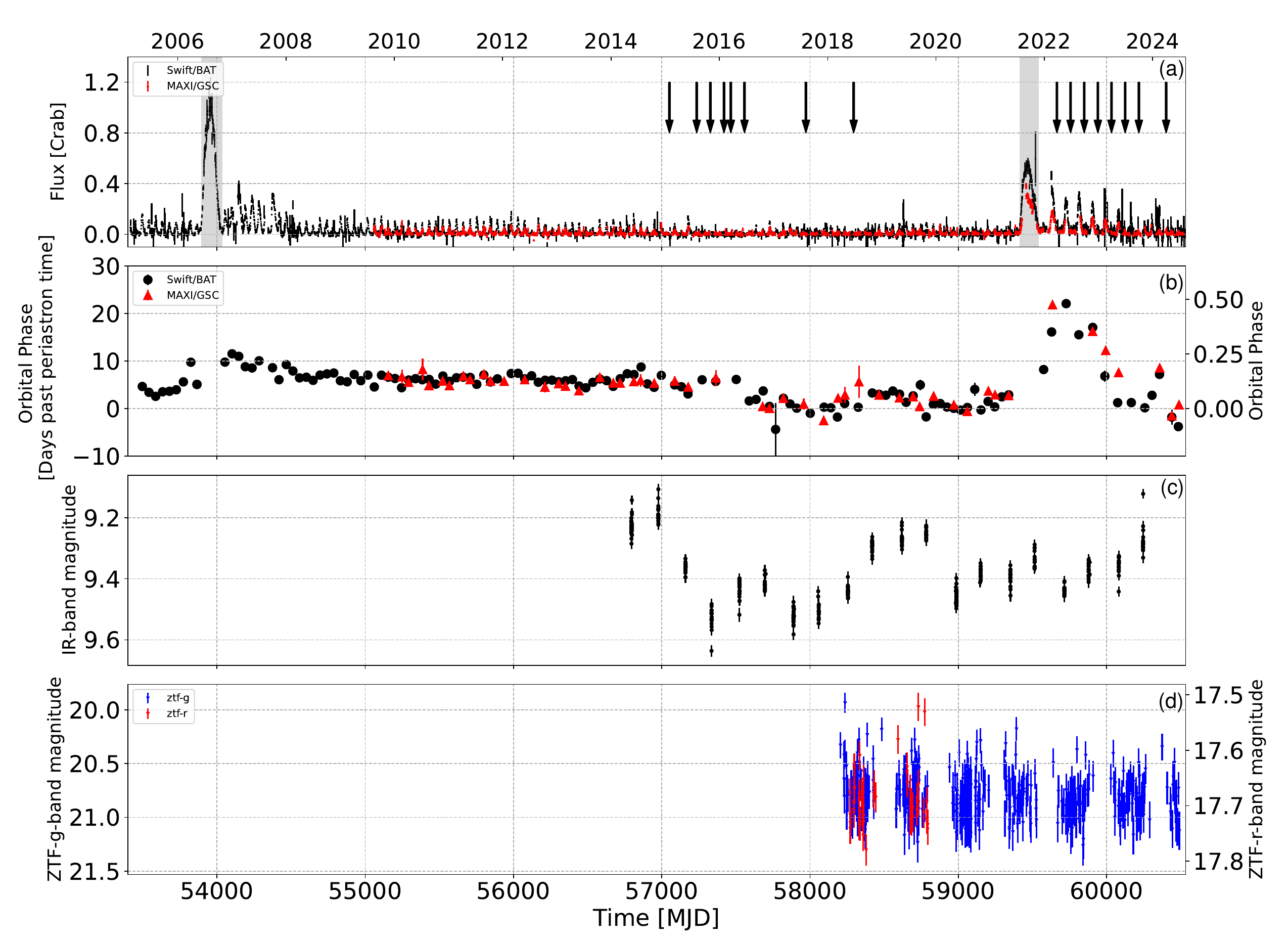}
    \caption{Panel a: the long-term {\it Swift}/BAT lightcurve of EXO 2030+375 in the energy range of 15-50\,keV, and MAXI/GSC in the energy range of 2-20\,keV.
    The shaded regions indicate giant outbursts in 2006 and 2021.
    Black arrows indicate the missing Type-I outbursts.
    %{\color{blue} (JL: maybe try a log scale for the y axis? I do not know if this will make improvements, but we can try...)}
    Panel b: orbital phases of Type-I outburst peaks, where the periastron time was estimated using the ephemeris adopted from \citet{Wilson2008}.
    %{\color{red} (JL: here I have a few comments: 1, enlarge the font size of labels; the default setting of all ploting software is too small for a hardcopy. 2, I wonder why there are no errors for orbital phases? Because they are too small to be recognized? Or you just missed them? Note: In science, a number is ONLY meaningful if we know its uncertainties. 3, I suggest adding another X-axis presenting years. For an example, see Figure 1 in https://arxiv.org/pdf/2406.12155.
    %4, why are there no MAXI data presenting here, but only shown in Figure 2. 
    %5, when saving a plot, vector graphics formats, such as *.eps, *.ps, *.pdf, are highly recommended. Unlike *.png, they can still keep a sufficient resolution when you zoom-in a plot, which is required for high-quality publications.)}
    Panel c: long-term infrared lightcurve of NEOWISE at 3.4$\mu$m.
    Panel d: long-term ZTF lightcurves of g (blue) and r (red) bands.
    }
    \label{fig1}
\end{figure*}

Because EXO 2030+375 exhibits regular and steady activities with outbursts at almost every approach to the periastron passage (although occasionally some outbursts are missing as reported by \citet[][]{Wilson2002, Laplace2017}).
It has been continuously observed by {\it Swift}/BAT since 2006 and {\it MAXI} since 2009.
We show long-term {\textit Swift}/BAT and {\it MAXI} lightcurves in Figure~\ref{fig1}, where the fluxes were re-scaled to Crab units for the sake of comparison, i.e., 1\, Crab=0.22\, BAT\,\,$\rm counts\ s^{-1}\ cm^{-2}$ at 15-50\,keV and 1\, Crab=3.6\, MAXI\,\,$\rm counts\ s^{-1}\ cm^{-2}$ at 2-20\,keV.
Tens of regular Type-I outbursts and Type-II outbursts in 2006 and 2021 can be clearly identified in both soft and hard X-ray bands. 
%{\color{red} (Here maybe we use arrows to hightlight where type-II outbursts are)}
%According to light curve, we used Gaussian model(Eq.\ref{gaussian}), asymmetric Gaussian model(Eq.\ref{assymetric Gaussian}) to fit each of Type \text{I} outbursts and for some outbursts showing double peak were fitted with a double Gaussian model after 2006.

\begin{figure*}
    \centering
    \includegraphics[width=1\linewidth]{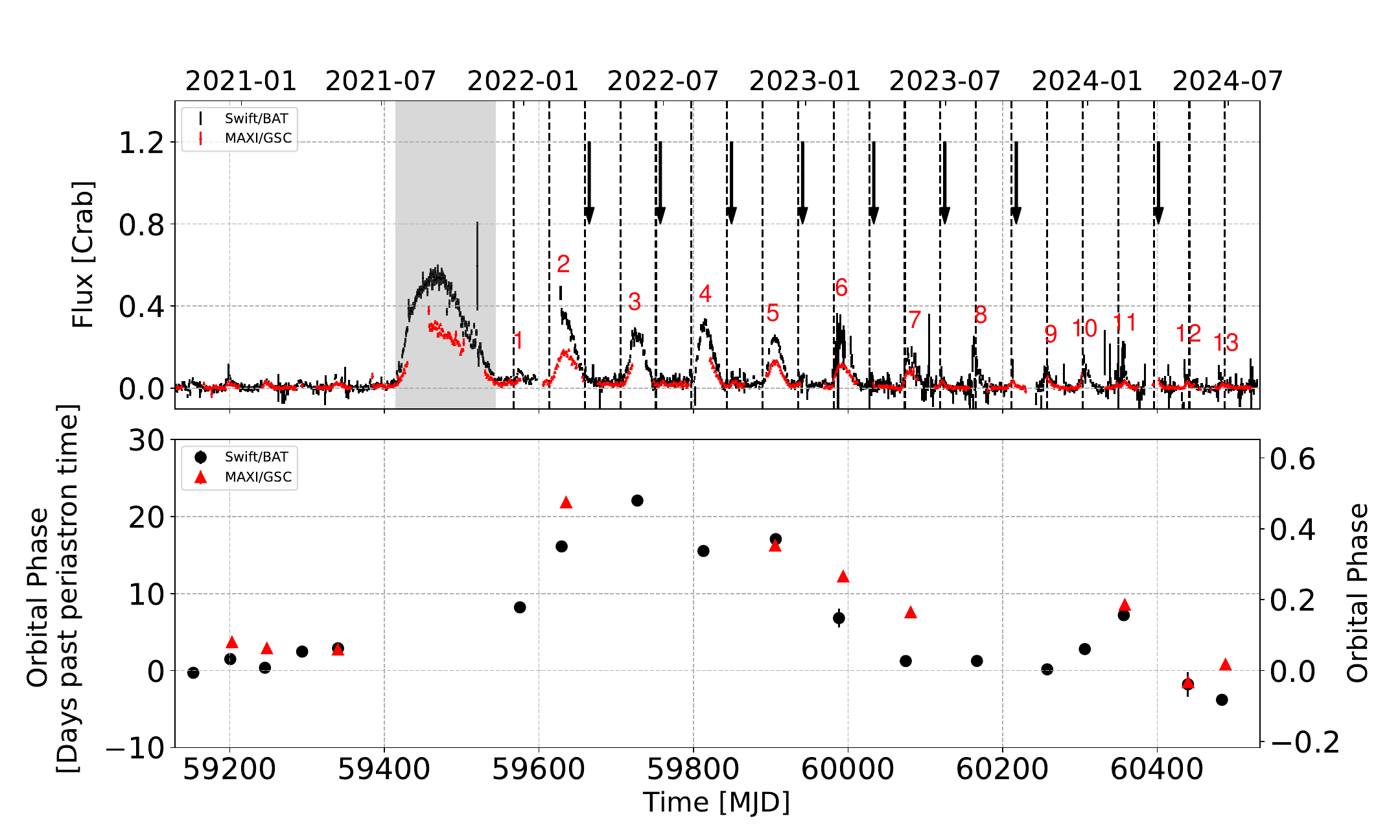}
    \caption{Similar to panels a and b in Figure~\ref{fig1}, but with a zoomed-in period close to the 2021 giant outburst.
    For clarity, vertical lines present periastron passages, and the arrows marks the missing outbursts which almost take place every two orbits.
    The red numbers denote the outbursts that occurred after the 2021 giant outburst.
    }
    \label{fig2}
\end{figure*}

To determine when Type-I bursts occurred, we applied both a Gaussian model ($F_{\text {Gau}}(t)=K\, e^\frac{-(t-t_{\rm peak})^2}{2 \sigma^2}$) and an asymmetric Gaussian 
($
F_{\rm aGau}(t) = K\, \left\{\begin{array}{l}
{\rm e}^\frac{-(t-t_{\rm peak})^2}{2 \sigma_{\rm 1}^2},\ {\rm when}\ t\leq t_{\rm peak}\\
{\rm e}^\frac{-(t-t_{\rm peak})^2}{2 \sigma_{\rm 2}^2},\ {\rm when}\ t>t_{\rm peak}
\end{array}\right.
$)
to describe each outburst and estimate their peak times\footnote{
Other models (e.g., a Lorentzian model and a double Gaussian model) have been used in literature, while as reported by \citet{Laplace2017}, they lead to very similar results compared to the Gaussian model.}.
Several outbursts with significant missing data (due to observational gaps) were excluded from our sample.
{In the end, we adopted the results from the model with a lower Akaike Information Criterion (AIC=$\chi ^{2}+2k+(2k^{2}+2k)/(n-k-1)$) value, where $\chi^2$ is the goodness-of-fit and $k$ is the number of free parameters and $n$ is the number of observation points.}
Since Type-I outbursts in EXO 2030+375 are consistently stable and occur near periastron passages, we were able to calculate the time difference between outburst peaks and the expected periastron passages using the ephemeris from \citet{Wilson2008} (Figure~\ref{fig1} and~\ref{fig2}).
Previously, similar morphological studies have been performed by different authors \citep{Wilson2002, Wilson2005, Baykal2008, Laplace2017}.
In this paper, we focus on the behavior related to the 2021 giant outburst and its comparison with the 2006 giant outburst.
Before 2006, peaks of Type-I outbursts occurred 5-6\, days after periastron, which suddenly jumped to about 14\, days after periastron associated with the 2006 giant outburst, followed by a slow recovery to 5-6 days.
After that, orbital phases of Type-I outbursts remained stable until MJD 57000 and then gradually shifted to near the periastron around MJD 58000, which had modulations afterwards.
Before the 2021 giant outburst, Type-I outbursts occurred at about 3\, days after periastron, similar to what was observed during the 2006 giant outburst.
During the 2021 event, the timing of the Type-I outburst peaks significantly delayed, shifting to as much as 20 days after periastron. 
Then , the phase shift recovers gradually in the next $\sim$500 days.
Another interesting thing is that several Type-I outbursts were missing after the giant outburst, resulting in Type-I outbursts occurring every two orbital cycles.
Here, we defined a missing outburst if the best-fitting $K$ is smaller than 3 times its uncertainty, i.e., $K/\sigma_{k} < 3$.  

The infrared light curve (panel c in Figure~\ref{fig1}) also presents long-term variations, suggesting the changes in the Be disk.
However, it does not seem to be associated with the giant outburst.
On the other hand, this source's optical magnitude is relatively faint, and no significant changes are found in the ZTF-g and ZTF-r light curves.

\section{Discussion and conclusion} \label{sec:discussion}
Type-I outbursts are thought to occur when the neutron star approaches the decretion disk of the Be star, provided that the outer radius of the Be disk extends to the Roche-lobe radius \citep[e.g.,][]{Okazaki2001, Okazaki2013}.
In EXO 2030+375, these outbursts appear consistently, which is unique among X-ray pulsars, potentially due to the presence of an unusually large Be disk \citep{Monageng2017}.
If the Be disk is uniform and stable, we would expect that the peaks of Type-I bursts appear at an unchanged orbital phase, which contradicts the observations.
For example, in 1995, EXO 2030+375 showed an orbital phase shift of Type-I outbursts from peaking at 6\, days after periastron to before periastron, followed by a slow recovery.
Such behaviour is very similar to what we observed in 2016 and was explained by a global one-armed oscillation propagating in the Be disk \citep{Wilson2002}.
Conversely, orbital phase shifts were also observed in 2006 and 2021 during giant outbursts, but with an opposite shifting direction.
It remains unclear whether these shifts originate from the same physical process, though this interpretation might be plausible since both Type-I and -II outbursts are triggered by accretion from the donor star and are ultimately determined by the properties of the Be disk.

\begin{figure}[H]
    \centering
    \includegraphics[width=0.8\linewidth]{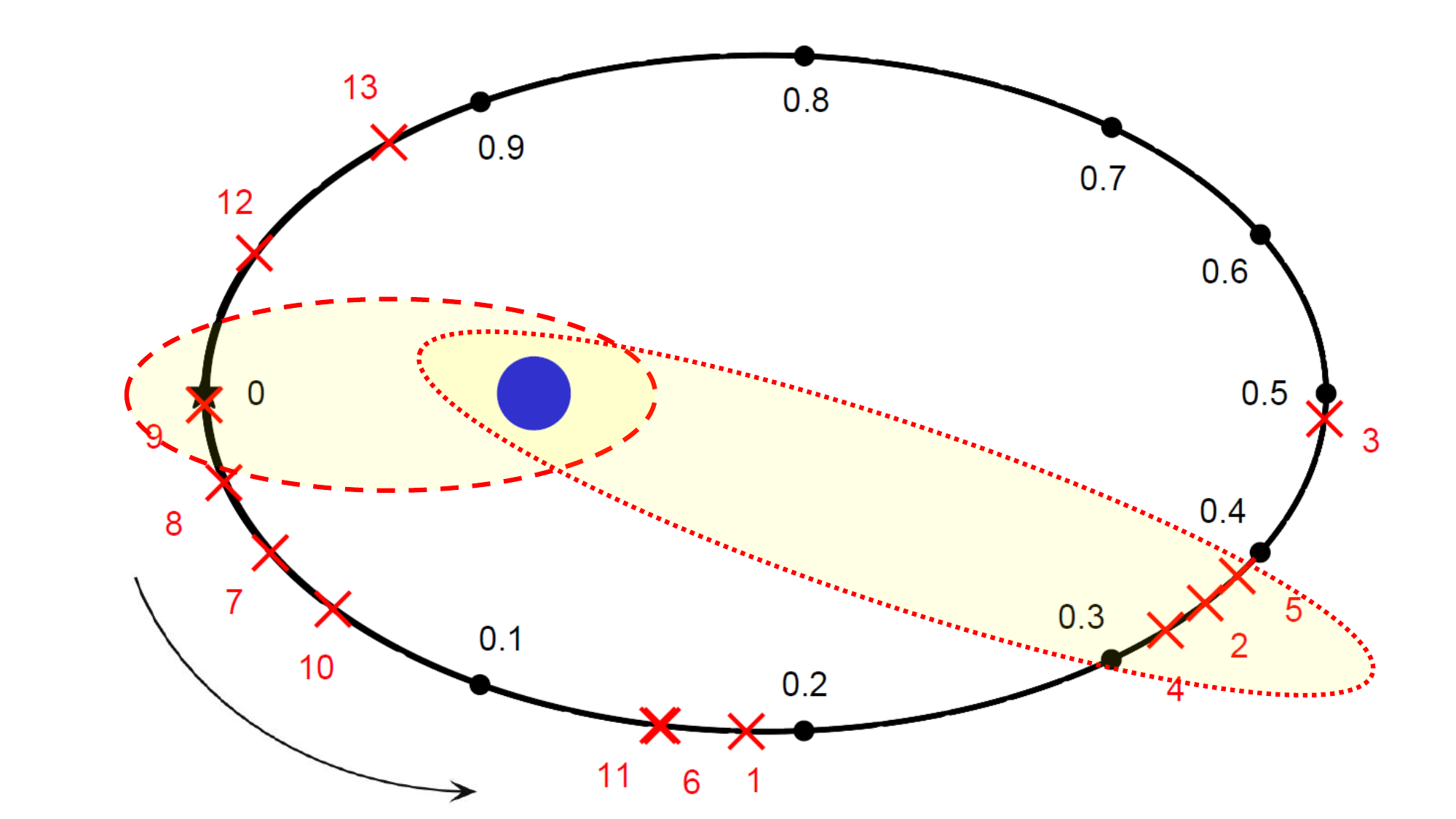}
    \caption{The orbital phase diagram of EXO 2030+375.
    The red numbers and crosses indicate orbital phases of Type-I outburst peaks after the 2021 giant outburst.
    The phase variation may reflect the changes of the Be circumstellar disk.
    Before giant outbursts, the neutron star passes through the Be disk (dashed line) near periastron.
    While during giant outbursts the Be disk might be significantly distorted and highly eccentric (dotted line), resulting in subsequent Type-I outbursts far away from the periastron.
    }
    \label{ObitalFigure}
\end{figure}

The formation of Type-II outbursts is likely due to precessing warped Be disk driven by radiation or tidal effects \citep{Negueruela2001, Martin2011}.
%A commonly accepted model for the Be disk is the viscous decretion disk, which reverses material transport from an accretion disk and explains that the inner disk near the star moves material outward to the outer disk in Be disks.
%This model successfully explains most of the characteristics of Be stars and predicts the truncation radius of Be disks. 
%It predicts that the truncation radius of EXO 2030+375 should be located at the 4:1 resonance radius (\citet{Okazaki2001}).
%However, Figure 5 of \cite{Monageng2017} shows that data from 1986 to 2007 indicate that the radius of Be disk is greater than 4 times resonance radius and even extends beyond periastron passage.
%And, comparisons of radius with outburst relationships from other sources indicate that outburst occur regardless of whether disk is relatively large or small. Therefore, Be disk size does not appear to correlate with Type \text {II} outbursts.
%According to existing data, there is no evidence of truncation in Be disk of EXO 2030+375.
%\cite{Negueruela2001} proposes that Type \text {II} outbursts are initiated When the disk around a Be star is truncated by the gravitational influence of a neutron star, its density gradually increases. 
%Once the disk achieves sufficient optical thickness, radiation pressure induces disk distortion. 
In this scenario, the observed phase shift would be expected due to the precession of the Be disk.
\citet{Martin2014a} conducted representative simulations using the smoothed particle hydrodynamics code to study the accretion rate over 50 orbits, where both Type-I and giant outbursts are present.
We extracted outburst peak times from their simulations and calculated the corresponding phase shifts (shown in Figure~\ref{Martin_simulations}).
%{\color{red} (JL: I need the new plot you made before by comparing Martin2014a's simulations and our observations)}.
The dramatic phase jump associated with the giant outburst aligns with our observation, although the gradual phase shifting during other periods is usually not evident in the actual data.
%Post-distortion, the disk  precession, where it tilts and rotates periodically. 
%Such disk distortion and precession can lead to overflow of material from the disk's outer regions towards the neutron star, triggering Type \text {II} X-ray outburst.
We note several Type-I outbursts after the 2021 giant outburst occurred at phases significantly distant from the periastron.
This is possible only if the Be disk is highly eccentric. 
%which is consistent with theoretical estimations \citep{Martin2014a}.
In the case of a warped and eccentric disk, the outer Be disk can extend further and reach the Roche lobe even near the apastron due to decreased Lindblad resonances \citep{Lubow2015}.
A possible picture is illustrated in Figure~\ref{ObitalFigure},
%{\color{blue}( Here we need a schematic. I am not good at 3D drawings, but I can try ...)},
which could explain the large shifts in Type-I outbursts observed after the 2021 giant outburst.
If this is correct,  the timescale for the recovery of outburst shifts corresponds to the timescale of the Be disk precession, which is about 500 days. 
In addition, the several missing bursts following the giant outburst could be explained as a faster and chaotic variation of the disk caused by the interactions with the neutron star, although detailed processes remain unknown.
%\textcolor{blue}{This scenario can indeed explain the irregular occurrence of Type-I outbursts after the 2021 giant outburst.
%The precession of the highly eccentric circumstellar disk leads to neutron star interactions occurring at intervals longer than the orbital period, resulting in a less periodic sequence of outbursts. 
%This chaotic variation of the disk shape and orientation during this time can be interpreted as a temporary disruption in the usual periodicity of Type-I outbursts.}

\begin{figure}[H]
    \centering
    \includegraphics[width=0.7\linewidth]{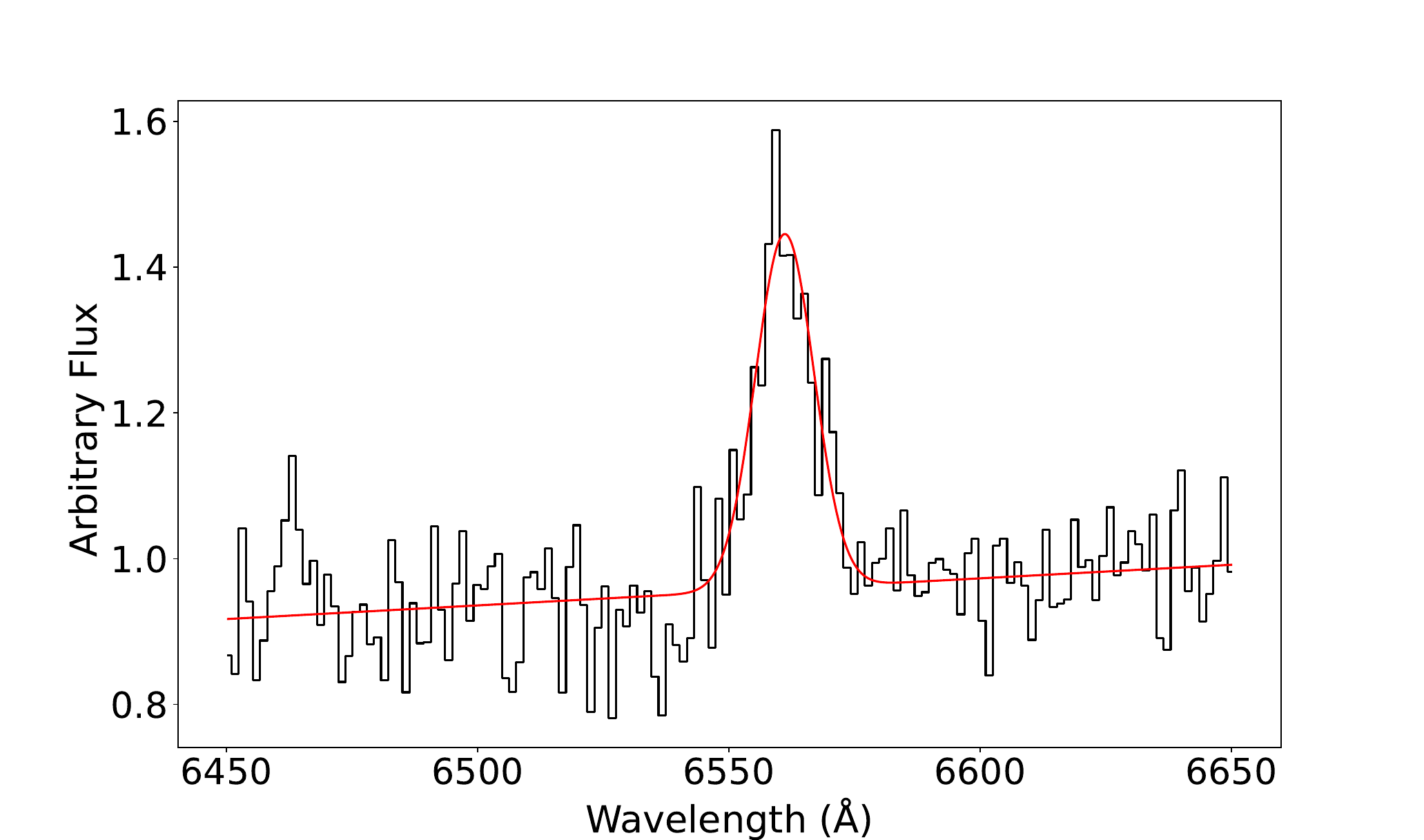}
    \caption{An optical spectroscopy of EXO 2030+375 observed with 2.4-m Lijiang telescope on November 8, 2021.
    The $H_{\rm \alpha}$ line can be fitted with a Gaussian model and its equivalent width is -6.62\,\AA.
    %{\color{red} JL: 1) the unit $\rm \AA$ should not be italic. 2) a larger font size }   
    }
    \label{halpha}
\end{figure}

\begin{figure}[H]
    \centering
    \includegraphics[width=0.7\linewidth]{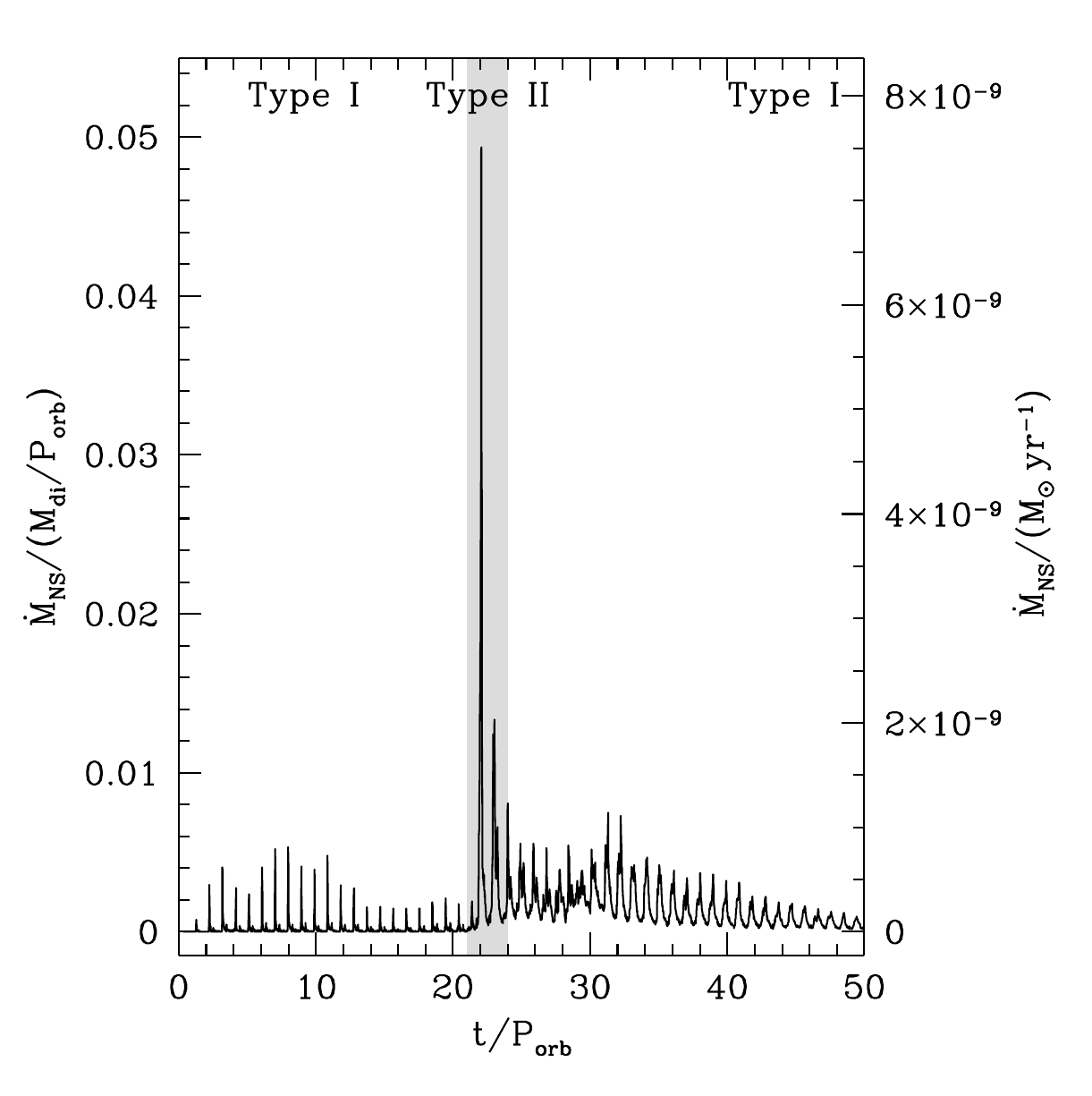}
    \includegraphics[width=0.7\linewidth]{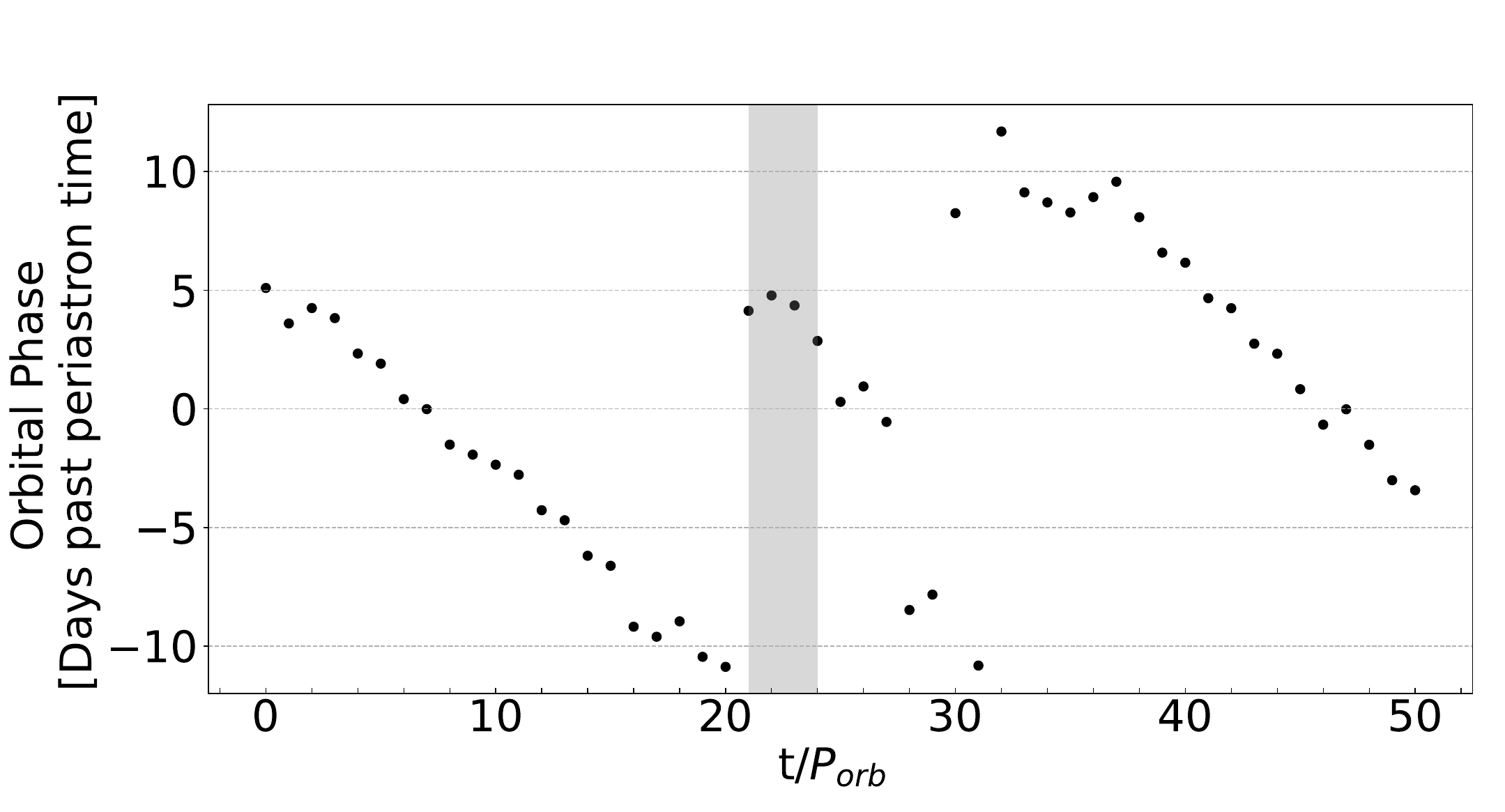}
    \caption{ Upper panel: simulations of the mass transfer rate onto the neutron star during Type-I and Type-II outbursts performed by \citet{Martin2014a} (adopted from Figure~4 in their paper), where the shaded region highlights the Type-II outburst.   
    Lower panel: evolution of orbital phases of Type-I outburst peaks according to simulations.
%    The mass transfer rate on to the neutron star, measured at the accretion radius of 1 $R_{\odot }$, is depicted in Figure. 4 from  \cite{Martin2014a}.
    }
    \label{Martin_simulations}
\end{figure}

In observations, the size and structure of the Be disk can be measured by monitoring the $H_\alpha$ emission line profile because its equivalent width (EW) serves as an indicator of the Be disk's size, and its shape reflects the disk's inclination \citep{Hummel1994, Liu2022, Liu2024}.
We conducted optical spectroscopy using a 2.4-m Lijiang telescope on November 8, 2021 (shown in Figure~\ref{halpha}) and obtained a $H_\alpha$ EW of -6.62\,\AA.
This value is similar to those measured in earlier years \citep{Wilson2002, Baykal2008, Furst2017}.
%{\color{red} (is this statement correct?; need citations about previous observations)}
%{\color{blue} {(Here we also need to show and describe the shape of the Halpha line (i.e., winebottle-type? single? double? red shoulder?) and compare with previous studies (also need citations), in order to test if the inclination angle changes.)}}
After the giant outburst, we observed no significant variability in the infrared and optical photometry. 
%One possible explanation is that, although the disk was significantly distorted, it may have remained sufficiently dense and intact to fuel the Type-I outbursts, which continued to occur in a less periodic manner.
These phenomena suggest that the amount of matter contributing to optical and infrared emissions in the Be disk remains stable, even though its structure may have been changed substantially.
Another possibility is that the optical and infrared emissions mainly originate from the outer part of a very large disk (larger than the binary orbit), where the distortion is insignificant. 
Up to now, there have been only four dramatic outburst phase shifts. We summarize their characteristics as follows:
1) the events with a negative phase shift (i.e., in 1995 and 2016) do not coincide with giant outbursts.
2) on the other hand, the events associated with giant outbursts exhibit a positive phase shift.
3) these two types of events appear to alternate, occurring approximately every 5-10\, years. 
4) after these events, the phase of Type-I outbursts typically shows a gradual recovery, although it may not always return to its original value.
\citet{Laplace2017} suggests that these phenomena might be attributed to a periodicity of approximately 10\, years, potentially caused by the Kozai-Lidov effect, i.e., a long-term periodic exchange between the eccentricity and the inclination of the Be disk \citep[see, e.g.,][]{Martin2019}.
If this is correct, we predict that the next significant orbital phase shift will be negative and occur between 2026 and 2031 without being associated with a giant outburst.

% \begin{figure}
%     \centering
%     \includegraphics[width=0.7\linewidth]{Martin2014_OP.pdf}
%     \caption{Orbital phase of outburst peaks, with timing derived from Figure. 4 from \cite{Martin2014a}.
%     }
%     \label{Martin2014_OP}
% \end{figure}

\begin{acknowledgments}
We thank Prof. Pu Du for providing the optical data and discussions.
This work is supported by the National Natural Science Foundation of China under grants No. 12173103 and 12261141691. 
\end{acknowledgments}

\end{document}